**Grading by Category: A simple method for providing students with meaningful feedback on exams in large courses.**

Cassandra Paul,[i] Wendell H. Potter[ii] & Brenda Weiss

Many instructors choose to assess their students using open-ended written exam items that require students to show their understanding of physics by solving a problem and/or explaining a concept. Grading these items is fairly time consuming, and in large courses time constraints prohibit providing significant individualized feedback on students' exams. Instructors typically cross out areas of the response that are incorrect and write the total points awarded or subtracted. Sometimes, instructors will also write a word or two to indicate the error. The experience of many instructors, however, is that with this level of feedback, students are not motivated to take advantage of the opportunity to engage in the feedback process to enhance their learning; their involvement stops with answering the question, "What grade did I get?" As illustrated in Carol Evans's recent review article, numerous education researchers have noted that providing feedback on exams can contribute to closing the gap between what students demonstrate they have learned and the performance we expect.[1] This paper describes a grading method that provides greater individualized feedback, clearly communicates to students expected performance levels, takes no more time than traditional grading methods for open-ended responses, and seems to encourage more students to take advantage of the feedback provided.

The "Grading by Category" (GBC) method was developed at the University of California, Davis (UC Davis) for use in 300+ student active-engagement introductory physics courses.[2] It has been in use for over 15 years, and is currently being used by dozens of instructors at UC Davis, and other institutions.

While GBC shares some similarities with rubric grading, it differs in three important ways:

- The categories emerge from the students' actual responses instead of being predetermined.
- All student responses are categorized prior to assigning a numeric score to each category.
- Each category focuses on the errors made by the students, not on the number of steps they complete correctly.

At UC Davis the GBC process is used to assess exam items that require students to reveal their thinking, often by drawing diagrams or writing explanations as well as those requiring calculations. After an exam the instructor views 20-30 student responses to each exam item to get an idea of the common errors. Instructors often find it helpful to place sticky notes on exams summarizing student errors as they sort solutions into different piles. As the common errors become evident, instructors define a category for each error or set of similar errors.

---

[i] San José State University, San José, CA
[ii] University of California – Davis, Davis, CA

Each category is given a unique symbol (usually a letter to facilitate grade entry). Only this symbol is written on the student's exam. Each exam problem or question is assigned its own symbol. Thus, an exam with three items would be marked with three symbols; e.g., XLP. At this stage in the process, numerical scores have not yet been determined, and when they are, they are not added to the student papers.

The symbols for each student exam are then entered into a spreadsheet or database, which associates the assigned grade point values with each category, which the instructor has assigned based on her/his criteria.[3] The instructor posts the categories with their associated grade values online and returns the students' exams with only the symbols on the top, so that the students are required to actually look at the Category Definition page in order to determine their grade.

We show below an example of a recent exam item along with the categories that were developed from the student responses. The categories, the grade points assigned to that category, and the percentage of students assigned to that category were posted online in the form that you see here. Students calculated their grades using this information along with the categories marked on their exams.

**Exam Question:**

There are two different equations listed on the equation sheet for acceleration under "linear/angular relation"; a = $\alpha$ r and a = $v^2$/r.  <u>Draw a picture of these two vectors, AND explain how they are related to each other</u>.  Are they the same? If so, why?  Are they different?  If so, how?

**Category Definitions and Grade Point Values:**

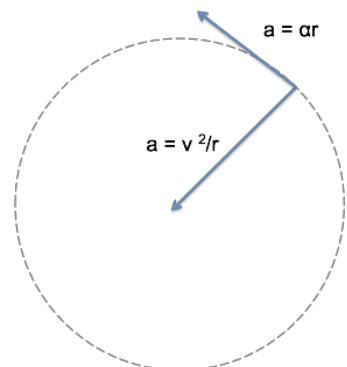

**Q (4.0):** Complete response: The two vectors are perpendicular to each other. a=$v^2$/r is the centripetal acceleration and is pointed towards the center of the circle an object is rotating about. a=$\alpha$ r is the tangential acceleration and is pointed tangent to the circle that the object is traveling in. (27%)

**M (3.5):** Same as Q, but mixed up the vectors. (6%)

**L (3.2):** Ideas were mostly complete and the same as in Q, but the pictures were incorrect or unclear.  Must discuss the tangential and centripetal acceleration to be in this category. (10%)

**C (2.5):** Some correct information was given regarding these quantities, but drawings were incorrect or unclear. The definitions of the vectors (as indicated in Q) were not given, and the fact that the vectors are perpendicular to each other was not indicated. (6%)

**P (2.0):** Argued that one of the equations represented an angular acceleration and one represented a linear one. (14%)

**S (1.0):** Argued that the accelerations equations were the same, but you just use them in different instances. (22%)

**N (0.5):** Some thoughts shown, but did not successfully complete the question. (10%)

**Z (0.0):** Blank or essentially blank (4%)

The major benefit of the GBC method is that students are afforded much more individualized feedback. When students view the categories online, they locate their category and determine the relative value of their particular response. They can see what error(s) they made, and how many other students made the same error. Also by comparing their response with the correct solution, they have the opportunity to make sense of why their response did not meet the instructor's expectation for the learning objectives assessed with this item.

Some of the additional benefits that the GBC method affords are briefly discussed in the remainder of this article.

GBC allows students to focus on their most significant errors. Categories are based on the most significant error made. If two student responses to a question on motion for example, show a similar violation of conservation of energy, their solutions would be placed in the same category, even if one included an additional minor mathematical error. (The instructor would indicate this by including a clause in the category stating that 'the solution may contain a minor mathematical error.') In contrast, a student response with a correct solution except for a minor mathematical error would not be placed in the same category as the student who solved it completely correctly.[4] This places the emphasis on the major error allowing students to focus on the more important misunderstandings or lack of understanding, rather than focusing on minor details and or mistakes.

GBC allows the instructor to wait until all the solutions are seen before making a final judgment on the value of a particular error. For example, note that students who made errors consistent with the category "M" switched the definitions when they described the vectors. Initially this was considered indicative of a major conceptual error; however, after further thought the instructor decided that the solutions provided enough evidence of understanding to indicate the students just accidentally switched the labels in the test situation. Whether or not you agree with the instructor's choice is irrelevant. The point is that this method allowed the instructor to give more value to solutions of this type AFTER the grading was completed, but before scores were distributed. Instructors sometimes figure out halfway through grading an exam that there is a reason students are making an odd error (such as poor item wording or a typo) and the GBC method saves the grader from choosing between re-grading the whole set, and giving some students a lower score than they deserve. This is especially useful to instructors who have graders who can do the actual sorting into categories, but retain the final say on the students' scores.

GBC provides the instructor with a quantitative report of student errors. With GBC, the grade spreadsheet can easily reveal the number of students who make the same error, which gives the instructor information regarding class-wide misconceptions or difficulties. Using traditional grading the best the instructor could do is determine the average score for each problem, and this is only if each numerical grade is entered separately for each problem. Information on common errors can help the instructor address areas where students are struggling with this class and future classes.

GBC affords a simple re-grade process. If students feel that their response was incorrectly categorized, they can request a re-grade. The student fills out a form[5] stating the category they were placed in and what category they believe they should be placed in, or why they believe that the particular category should receive more points. They must also solve the problem correctly. In the authors' experience students are less likely to challenge a grade with this method, but whether this is because they understand why they received the score that they did, or because effort is required of them to prove their understanding has yet to be investigated. The re-grade process relieves some of the agony of assigning partial credit because if the instructor knows the students can request re-grades if they understand the material, but their solution is difficult to follow, instructors are less likely to spend an inordinate amount of time trying to decipher student work. This puts the 'burden of proof' on the students to show their understanding, instead of the instructor, which Henderson suggests can encourage students to explain their reasoning in future exam situations.[6]

Student response to GBC is positive overall. The only complaint students have about GBC is that they don't immediately have access to their numerical score. When asked about the GBC method, most students will identify at least one reason they like it more than grading techniques used in their other science courses. Several typical responses are listed here:

*"I liked how you put the letters instead of numbers on the test so that we could get more feedback without all the red marks on the physical tests."*

*"I think in general the system is nice because not everyone really wants to show off their grades straight up to everyone."*

*"I felt that it … helps the students get a better grade then they … would have if you used a more traditional method. It also allows the teacher to get a better idea of where the students are having trouble; this can help you help us. "*

*"From the [categories], I was able to understand where I had made mistakes and it helped guide me in the right direction to solve the problem. The grading system made*

*me also look closer at the exam as opposed to looking at my grade or percentage and hiding it in my binder."*

GBC encourages students to contemplate their grade when they have time to think about it. Instructors often wish students would figure out what they did wrong, but they don't always create opportunities to do this. Exams are often handed back at the end of class when students are rushing out the door. Often students will look at their score and a subset will be satisfied with their score and never look at it again. Another subset will be disappointed and also not want to look at their exam again. Denying students the instant knowledge of their score allows them to contemplate it when they have they have the resources (time and categories) to make sense of what they did wrong. In this way GBC encourages students to study the correct solution, as well as their errors. The categories help students figure out how their particular solutions relate to the instructor's expectations, and thus help communicate the important features of the course.

We believe that assessment is the best way instructors can convey expectations to students, and that students continue to learn after the assessment is given. With the Grading by Category method we not only provide our students more feedback for improvement, but that feedback is matched and weighted to appropriately communicate the relative value of the skills and concepts they learn in our class.

For resources and further instruction on the Grading by Category method, please visit our website: http://www.sjsu.edu/people/cassandra.paul/gradingbycategory/

---

[1] Carol Evans, Making Sense of Assessment Feedback in Higher Education. Review of Educational Research, March 2013, Vol. 83, No. 1, pp. 70-120

[2] Wendell Potter, David Webb, Emily West, Cassandra Paul, Mark Bowen, Brenda Weiss, Sixteen years of Collaborative Learning through Active Sense-making in Physics (CLASP) at UC Davis arXiv:1205.6970 [physics.ed-ph]

[3] A template for such a spreadsheet is provided at our GBC website.

[4] This is not to say that categories can consist of only one error. In situations where more than one conceptual error can be identified a category can represent multiple errors. For example, a category could represent a violation of conservation of energy <u>and</u> an additional misconception regarding potential energy. See our website for more examples.

[5] An example of this form is downloadable at our GBC website.

[6] C. Henderson, E. Yerushalmi, V. Kuo, P. Heller, and K. Heller, Grading Student Problem Solutions: The Challenge of Sending a Consistent Message, Am. J. Phys. 72, 164 (2004).